\documentclass[%
 reprint,
nofootinbib,
nobibnotes,
bibnotes,
 amsmath,amssymb,
 aps,
]{revtex4-1}
\usepackage{MnSymbol}
\usepackage{natbib} 
\newcommand{\diamondnew}{%
            \mathrel{\raisebox{.1em}{%
         \reflectbox{\rotatebox[origin=c]{45}{$\boxtimes$}}}}}

\newcommand{\comment}[1]{}
\usepackage{tikz}
  \newlength\squareheight
\usepackage{wasysym} 
\usepackage{graphicx}
\usepackage{dcolumn}
\usepackage{bm}

\begin{document}

\preprint{Draft}

\title{Percolation in random graphs with higher-order clustering}

\author{Peter Mann$^{1,2,3}$}
\email{pm78@st-andrews.ac.uk}
\author{V. Anne Smith$^2$}%
\author{John B.O. Mitchell $^1$}
\author{Simon Dobson$^{3}$}
\affiliation{$^3$ School of Computer Science, University of St Andrews, St Andrews, Fife KY16 9SX, United Kingdom }
\affiliation{$^1$ School of Chemistry, University of St Andrews, St Andrews, Fife KY16 9ST, United Kingdom }
\affiliation{$^2$ School of Biology, University of St Andrews, St Andrews, Fife KY16 9TH, United Kingdom }

\date{\today}

\begin{abstract}
    Percolation theory can be used to describe the structural properties of complex networks using the generating function formulation. This mapping assumes that the network is locally tree-like and does not contain short-range loops between neighbours. In this paper we use the generating function formulation to examine clustered networks that contain simple cycles and cliques of any order. We use the natural generalisation to the Molloy-Reed criterion for these networks to describe their critical properties and derive an analytical description of the size of the giant component, providing solutions for Poisson and power-law networks. We find that networks comprising larger simple cycles behave increasingly more tree-like. Conversley, clustering comprised of larger cliques increasingly deviate from the tree-like solution, although the behaviour is strongly dependent on the degree-assortativity. 
\end{abstract}

\pacs{Valid PACS appear here}
\maketitle


\section{Introduction}
\label{sec:introduction}


The generating function formulation, developed by Newman, Strogatz and Watts~\cite{newman_strogatz_watts_2001} is an extremely powerful combinatorial technique that can be used to elucidate structural properties of complex networks; such as: the distribution of small components, giant connected component (GCC) size, average degree, average path lengths, the onset of the phase behaviour, and so on. Typical network analysis assumes that the network is \emph{locally tree-like}: the number of short cycles is assumed to vanish as the network becomes infinitely large. However, it is well known that most real networks, including social and biological networks, contain shortcuts known as \emph{clustering}. Clustering occurs when a single node can be reached along multiple paths through the network, breaking the locally tree-like assumption. 


Newman provided the first mapping of the bond percolation process on an undirected configuration model network to epidemic processes on locally tree-like complex networks~\cite{Newman2002SpreadOE}. Miller and Newman independently extended this mapping to consider triangular clustering in random configuration model networks~\cite{PhysRevE.80.020901, PhysRevLett.103.058701}. Triangular clustering in complex networks has also been extensively studied using a variety of other techniques~\cite{PhysRevX.4.041020,PhysRevE.68.026121,PhysRevLett.97.088701, miller_2009}. In these studies, it was found \cite{PhysRevLett.103.058701} that Poisson-clustered networks cause a decrease in the critical bond percolation point, whilst also reducing the maximum size of the GCC. Conversely, it has also been shown that $\delta$-type clustering \cite{PhysRevE.80.020901} can increase the critical point whilst not always reducing the size of the GCC. These dichotomous results are a manifestation of the difficulty in experimental design when examining clustering; often, many other network properties (for example the degree assortativity) simultaneously shape the emergent properties. 

The properties of larger cycles and the influence of higher-order clustering on the structural properties of networks has been addressed in the literature using the generating function formulation before~\cite{gleeson_2009,benson_gleich_leskovec_2016,fronczak_holyst_jedynak_sienkiewicz_2002,hebert-dufresne_noel_marceau_allard_dube_2010,PhysRevE.97.052306,allard_hebert-dufresne_young_dube_2015,karrer_newman_2010,PhysRevE.68.026121,PhysRevE.81.066114,PhysRevE.93.030302}. Additionally, the related message passing formulation has incredible ability to describe the effects of clustering on percolation \cite{newman_2019,cantwell_newman_2019,karrer_brian_newman_j._lenka_2014}. However, the precise study of clustering comprising weak cycles and cliques, whilst investigating degree assortativity, has not yet been performed in the literature. 

\begin{figure}[ht!]
\begin{center}
\includegraphics[width=0.36\textwidth]{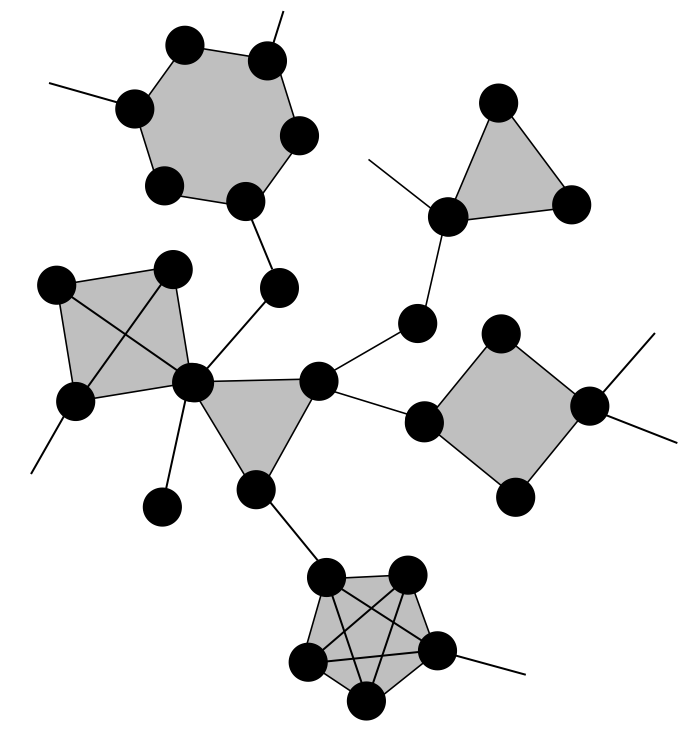}
\caption[network]{We consider networks containing locally tree-like edges, weakly clustered cycles and cliques of finite order. The method presented here can be extended to any cluster topology.
} \label{fig:Network}
\end{center}
\end{figure}

In this paper, we use the generating function formulation to consider networks containing additional cycles beyond triangles, and present analytical solutions for a variety of higher-order cluster topologies. For this purpose, we define \emph{strong} cluster topologies as cliques or maximal sub-graphs in the network. As we remove edges from the clique, the clustering is successively \emph{weakened} until, in its weakest form, only the outer skeleton of the topology remains and there are no shortcuts within the cycle; the cycle is a closed chain of nodes (see Fig~\ref{fig:Network}).

The paper is structured as follows: section~\ref{sec:framework} examines the percolation mapping for networks comprising clustered motifs of any order with the introduction of a $\tau$-dimensional joint probability distribution. Section~\ref{sec:howc} studies weak cycles of arbitrary order while in section~\ref{sec:HOCLIQUES} we derive a formula describing the $\tau$-clique. In section~\ref{sec:Percolation threshold} we examine the percolation threshold by formulating a generalised Molloy-Reed criterion for clustered graphs and compute the average component size and its distribution. We then study a joint Poisson degree distribution and clustered scale-free networks with exponential degree cut-off.

To examine the percolation properties of these networks, we introduce an approximate method to describe the complex correlations that arise in clustered motifs. The literature currently has exact methods to do this; however, they are either a recursive, numerical method \cite{allard_hebert-dufresne_young_dube_2015}, or an exponentially slow exhaustive enumeration of states for a given motif \cite{karrer_newman_2010}. Our method offers an equation-based framework built on the logic of \cite{PhysRevE.80.020901}. This method replicates the emergent properties of random configuration model networks with high accuracy to experimental bond percolation as shown in section \ref{sec:exactness}. Further, since it is based on the enumeration of pathways through a cycle, it can be used to examine the specific flow avenues through a network motif and identify, for example, super-spreaders and infection-conduits in epidemiology. 

All experiments are conducted according to the generalised configuration model described in \cite{karrer_newman_2010}. 

\section{Theoretical Framework}
\label{sec:framework}

In this section we review the generating function formulation \cite{newman_strogatz_watts_2001, karrer_newman_2010, miller_2009,PhysRevLett.103.058701} to consider clustered networks comprised of an arbitrary number of topological cycles. The underlying philosophy of this framework is that the degree of a node can be partitioned into sub-degrees that correspond to the involvement of a node in pre-defined topological cycles. For instance, the generalised degree, $\bm {k_\tau}=(k_\bot,k_{\Delta},k_{\square},\dots)$, of a node that has zero tree-like edges yet is a member of a triangle, two squares and three pentagons would be $\bm {k_\tau}=(0,1,2,3)$. The conventional degree distribution, $p(k)$, is the total number of edges of any kind emanating from a node. It is found using the Kronecker delta function $\delta_{i,j}$ as
\begin{equation}
    p(k)=\sum^\infty_{k_\bot=0}\cdots\sum^\infty_{k_\gamma=0}
    p({k_\bot,\dots,k_\gamma})\delta_{k,\sum k_{\tau\in\bm{\tau}}}
\end{equation}
\noindent where $\bm{\tau}$ is a vector of cluster topologies $\{\bot, \triangle, \square, \pentagon, \cdots \}$, $k_\tau$ is the degree of shape $\tau\in \bm \tau$ and $p(\bm{k_\tau})=p(k_\bot,\dots,k_\gamma)$ is the $\dim(\bm\tau)$ joint probability distribution of degrees. For instance, a node that is part of a two tree-like edges, a triangle and a square will have the following joint degree sequence $(k_\bot,k_\triangle,k_\square)=(2,1,1)$, while its overall degree is $k=6$. A network is described by its joint probability distribution of each node playing a certain role in a given cycle a particular number of times \cite{karrer_newman_2010}. The joint probability distribution is the probability of picking a node with a given joint degree sequence given all the degrees of the nodes in the network. The joint degree distribution can be generated using 
\begin{align}
     G_0(\bm{z})=& \sum_{k_\tau\in \bm{\tau}}^\infty p(\bm{k_\tau})\bm{z}^{k_{\bm{\tau}}}\label{eq:G0}
\end{align}
\noindent where $\bm z = \{ z_\bot, z_\triangle, z_\square, \dots \}$ and we have used the notation
\begin{equation}
     \sum_{k_\tau\in \bm{\tau}}^\infty = \sum^\infty_{k_\bot=0}\ \sum^\infty_{k_\triangle=0}\ \sum^\infty_{k_\square =0} \cdots 
\end{equation}
There are as many joint excess degree distributions as there are topologies in $\bm{\tau}$; they give the distribution of $k_\tau\in k_{\bm{\tau}}$ after following a random shape in $\bm{\tau}$ back to a node. For instance, $q_\bot(\bm {k_\tau})$ is the distribution of the number of tree-like edges, triangles, squares and so on, reached by following a randomly-chosen tree-like edge to a node. Similarly, $q_\triangle(\bm {k_\tau})$ is the distribution, arising by following a randomly-chosen triangle to a node. In general, for weak-cycles, the joint excess degree distribution is 
\begin{equation}
    q_\tau(\bm {k_\tau})=\frac{(k_\tau+1)p(k_{\bm\tau\backslash\{\tau\}}, k_{\tau}+1)}{\langle k_\tau\rangle}
\end{equation}
\noindent where the notation $\bm {\mathcal S}\backslash\{s\}$ excludes element $s$ from set $\bm {\mathcal S}$. We recover the tree-triangle model \cite{PhysRevE.80.020901, PhysRevLett.103.058701} from this formulation if we set $\bm \tau = \{\bot,\triangle\}$ and hence find 
\begin{align}
    q_\bot =& \frac{(k_\bot +1)p(k_\bot+1,k_\triangle)}{\langle k_\bot\rangle}\nonumber\\
    q_\triangle=&  \frac{(k_\triangle +1)p(k_\bot,k_\triangle+1)}{\langle k_\triangle\rangle}\nonumber
\end{align}

The joint excess degree sequence is generated as
\begin{align}
    G_{1,\tau}(\bm z) =&\ \sum^\infty_{k_\tau\in \bm\tau} q_\tau(\bm {k_\tau})\bm{z}^{k_{\bm{\tau}}}\nonumber\\
    =&\frac{1}{\langle k_\tau\rangle}\sum^\infty_{k_\tau\in \bm\tau}k_\tau p(\bm{k_\tau})\bm z^{k_{\bm\tau}\backslash\{k_\tau\}}z^{k_\tau-1}
\end{align}
\noindent and is also seen to be the partial derivative of Eq. \ref{eq:G0} with respect to $z_\tau$ divided by the expected number of $\tau$-cycles
\begin{equation}
    G_{1,\tau}(\bm z) = \frac 1{\langle k_\tau\rangle}  \frac{\partial G_0}{\partial z_\tau}\label{eq:Jacobian}
\end{equation}
which can also be written as 
\begin{equation}
    G_{1,\tau}(\bm z) = \frac{G_0^{'\tau}(\bm z)}{G_0^{'\tau}(\bm 1)}
\end{equation}
where $G_0^{'\tau}$ is the first derivative of $G_0(\bm z)$ with respect to $z_\tau$ and $\langle k_\tau\rangle=G_0^{'\tau}(\bm 1)$ is the average $\tau$-degree for a node in the network. We can also use this formula to compute the distribution of second nearest neighbours by following an edge in a $\tau$-cycle by composing the relevant $G_{1,\tau}(\bm z)$ with $G_0(\bm z)$ as 
\begin{equation}
    G_0(1,\dots ,  G_{1,\tau},\dots ,1)
\end{equation}

To evaluate the percolation properties of the network, we need to find the probability that the focal node does not become attached to the GCC through any of its neighbours. To achieve this we pick a node from the network at random and consider all combinations of events that could lead to its attachment. This probability depends on the type of cycle that the focal node connects to and for each topology $\tau\in \bm \tau$ we must write the probability that attachment does not occur through that cycle, $g_\tau$. The objective, therefore, is to find analytically a vector of $\bm {g_\tau} = \{g_\bot, g_\triangle, \dots\}$ for all topological cycles the node could be a part of. Each $g_\tau$ is a function of $u_\tau$ which is the probability that a node within a $\tau$-cycle is not attached to the GCC. In weak cycles and cliques, the probability $u_\tau$ is the same for each node site in the cycle; since, all node sites are identical. Once these probabilities have been found for the set of topologies $\bm \tau$ we average over the joint degree distribution to find the generating function, $G_0(g_\bot, g_\triangle, \dots,g_\gamma)$, for the probability of not being attached to the GCC as 
\begin{equation}
   \sum^\infty_{k_\bot=0}\sum^\infty_{k_\triangle=0}\cdots \sum^\infty_{k_\gamma =0}p(k_\bot,k_\triangle,\dots,k_\gamma)g_\bot^{k_\bot}g_\triangle^{k_\triangle}\cdots g_\gamma^{k_\gamma}
\end{equation}
or, using a condensed notation
\begin{equation}
    G_0(g_{\bm {\tau}}) = \sum^\infty_{k_\tau\in \bm {\tau}}p(\bm{ k_\tau})\bm{g_\tau}^{\bm {k_\tau}}\label{eq:uninfectedsize}
\end{equation}
where we compute each $u_{\mathbb \tau}$ element using the joint excess degree distribution generating function as a self-consistent equation
\begin{equation}
    G_{1,\tau}(g_{\bm {\tau}}) = \sum^\infty_{k_\tau\in \bm {\tau}}q_\tau(\bm{ k_\tau})\bm{g_\tau}^{\bm {k_\tau}}
\end{equation}
\noindent where $u_{\mathbb \tau} = G_{1,\bm\tau}(g_{\bm {\tau}})$ and solve using fixed point iteration. The sum converges with a solution $|g_\tau|\leq 1$. The size of the largest percolating cluster $S$ can then be calculated from one minus this quantity,
\begin{equation}
    S = 1 - G_0(g_{\bm {\tau}})\label{EQ:main}
\end{equation}
For instance, \cite{miller_2009, PhysRevLett.103.058701} the generating function for the tree-triangle model, $G_0(g_\bot,g_\triangle)$, can be recovered from this formulation by setting $\bm\tau=\{\bot,\triangle\}$ as 
\begin{equation}
    G_0(g_\bot,g_\triangle) = \sum_{k_\bot=0}^\infty\sum_{k_\triangle=0}^\infty p(k_\bot,k_\triangle)g_\bot^{k_\bot}g_\triangle^{k_\triangle}
\end{equation}
We then need to find the $g_\tau$ equations for a chosen vector of topologies that include weakly-connected and strongly-connected topologies of finite but arbitrary order. The remainder of this paper concerns the analytical description of $g_\tau$ for these cycles. 

\section{Higher-order weak clusters}
\label{sec:howc}

The bond percolation model considers the local environment of a randomly chosen node from the graph. Defining the edge occupancy probability to be $\phi$, we set the probability that a neighbour is not connected to the GCC to be $u_\bot$; i.e. that it belongs to a finite-sized component. As derived in \cite{Newman2002SpreadOE} for networks consisting entirely of tree-like edges, the probability that a particular chosen node does not become attached is the sum of the probabilities associated to all possible scenarios in which the neighbours of the focal node fail to attach it. Either the neighbour was not itself attached with probability $u_\bot$; or, it was attached but did not connect the focal node (i.e. failed to occupy their connecting edge) with probability $(1-u_\bot)(1-\phi)$. 

The presence of connections between branches of the neighbours of a node provides extra routes for the process to flow through which are ``unexpected'' and introduce additional complexity when considering the local environment of the focal node, correlating the probabilities of it remaining unattached to the GCC. 
\begin{figure}[ht!]
\begin{center}
\includegraphics[width=0.48\textwidth]{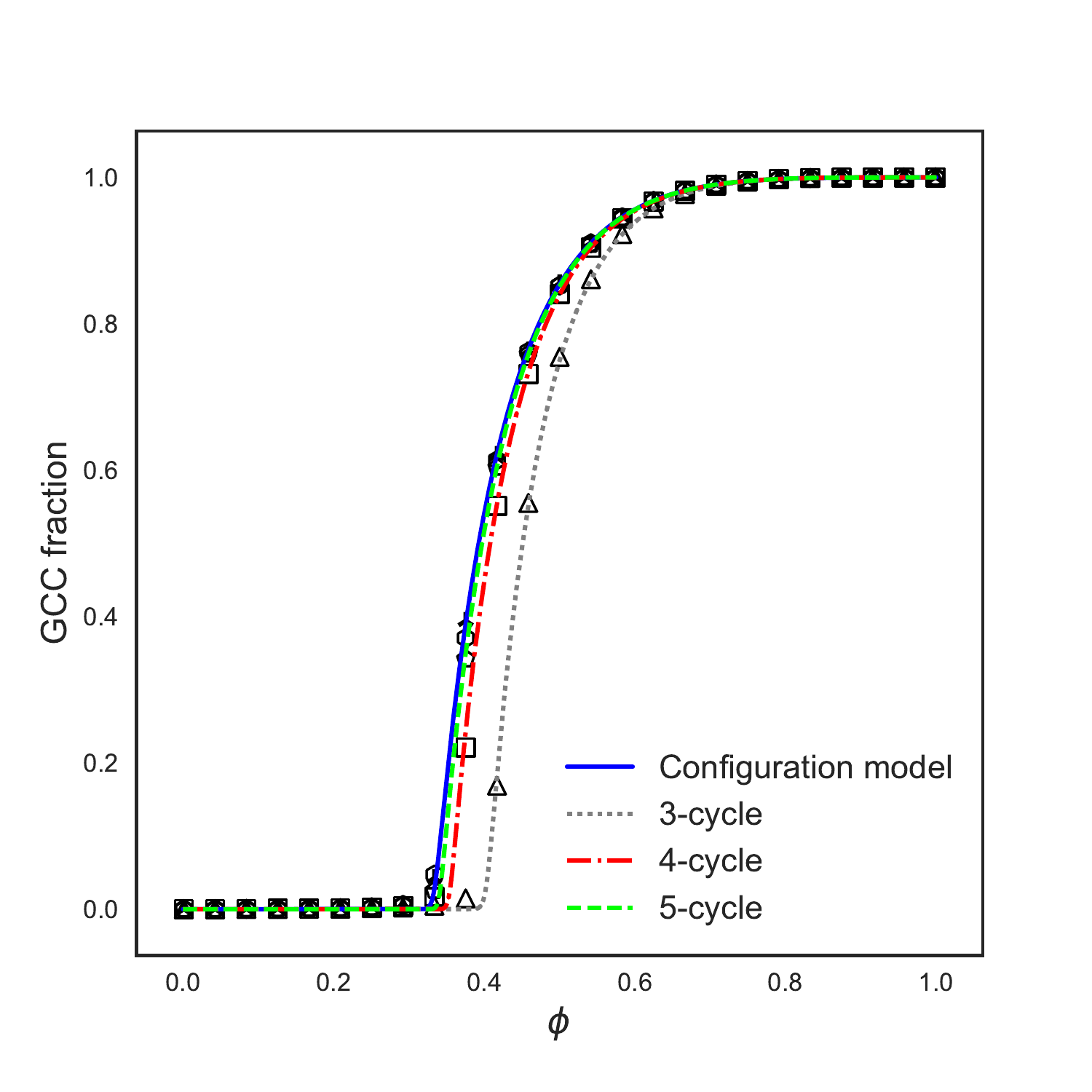}
\caption[weak]{
The fraction of the network occupied by the GCC for networks comprised of increasingly larger weak cycles. Curves are theoretical predictions whilst scatter points are simulation averages. 
} \label{fig:WeakOutbreaks}
\end{center}
\end{figure}
In this section, we consider clusters that contain higher-order cycles of finite but unspecified size. We refer to these as ``weak'' cycles. They contain no shortcuts back to the focal node or to other cluster nodes: they are simply closed chains of nodes.

We proceed by examining the chains of unconnected nodes between two contact nodes to a focal node. Weak cycles contain exactly two direct-contact nodes. The probability that they fail to attach the focal node to the giant component is again
\begin{equation}
    [u_{{\tau}} + (1-u_{{\tau}})(1-\phi )]^2
\end{equation}
\noindent However, they can still attach the focal node through the cluster if there is a chain of unconnected nodes, and bond occupation occurs at each step. We define there to be $\eta$ nodes in the topological shape, excluding the focal node. Then, for a contact node to occupy the focal node through the cluster there must be one contact node attached to the giant component $(1-u_{{\tau}})$, $\eta-1$ unconnected nodes in the shortcut, $u_{{\tau}}^{\eta-1}$, which each get connected sequentially from their neighbour $\phi ^\eta$ and direct connection failure, $1-\phi $. This results in 
\begin{equation}
    2(1-u_{{\tau}})(1-\phi )u_{{\tau}}^{\eta-1}\phi ^\eta
\end{equation}
\noindent with the multiplication by two accounting for the symmetrical counterpart to the shape.

Next, we must account for non-contact nodes around the body of the cycle. These must have a chain of contacts back to the focal node through which they connect the focal node to the percolating cluster. Simultaneously, the other direction through the chain must fail to connect the focal node (as this mode must be the successful one).

For the $i$th node in the body of the cycle, there must be $1\ldots i-1$ unconnected nodes back to the focal node and $i+1\ldots \eta$ nodes remaining in the structure. The probability of $i-1$ unconnected nodes and $i$ occupations from the occupied $i$th node is 
\begin{equation}
   (1-u_{{\tau}}) u_{{\tau}}^{i-1}\phi ^i
\end{equation}
\noindent The failure to connect through the remainder of the chain of $\eta-i$ nodes is a somewhat more complex process. There are two ways for the process to fail. Firstly, there \textit{is} a chain of unconnected nodes, the $i$th node \textit{does} connect to the GCC, but at some point along the chain it fails. Secondly, the process could reach an already attached node that also fails to connect the focal node. 

We introduce a new index, $l$ in the range $[i+1, \eta]$ that tracks the current node under consideration. One minus the probability that the $i$th node successfully connects the GCC to the focal node is
\begin{equation}
    1-u_{{\tau}}^{\eta-i}\phi ^{\eta-i+1}
\end{equation}
\noindent Similarly, the probability that all the nodes in the remainder of the cycle fail to connect to the GCC is 
\begin{equation}
    \prod^\eta_{l=i+1}\bigg[1-(1-u_{{\tau}})u_{{\tau}}^{\eta-l}\phi ^{\eta-l+1}\bigg]
\end{equation}
\noindent We can then construct the probability that the $i$th node in the cycle successfully connects to the focal node
\begin{align}
    &2(1-u_{{\tau}})u_{{\tau}}^{i-1}\phi ^i(1-u_{{\tau}}^{\eta-i}\phi ^{\eta-i+1})\nonumber\\
    &\times \prod^\eta_{l=i+1}\bigg[1-(1-u_{{\tau}})u_{{\tau}}^{\eta-l}\phi ^{\eta-l+1}\bigg]
\end{align}
\noindent where multiplication by two accounts for the symmetry of the cycle. To account for all body nodes we need to sum this from $2\ldots \eta-1$
\begin{align}
    &2\sum_{i=2}^{\eta-1}(1-u_{{\tau}})u_{{\tau}}^{i-1}\phi ^i(1-u_{{\tau}}^{\eta-i}\phi ^{\eta-i+1})\nonumber\\
    &\times \prod^\eta_{l=i+1}\bigg[1-(1-u_{{\tau}})u_{{\tau}}^{\eta-l}\phi ^{\eta-l+1}\bigg]
\end{align}
\noindent The vector of total probabilities that the focal node remains unattached when it belongs to a topological shortcut, $ {g_\tau}$, can then be constructed for any weak cycle topology 
\begin{align}
     {g_\tau} =\ &[u_{{\tau}} + (1-u_{{\tau}})(1-\phi )]^2\nonumber\\
    & -2(1-u_{{\tau}})(1-\phi )u_{{\tau}}^{\eta-1}\phi ^\eta\nonumber\\
    &-2\sum_{i=2}^{\eta-1}(1-u_{{\tau}})u_{{\tau}}^{i-1}\phi ^i(1-u_{{\tau}}^{\eta-i}\phi ^{\eta-i+1})\nonumber\\
    &\times \prod^\eta_{l=i+1}\bigg[1-(1-u_{{\tau}})u_{{\tau}}^{\eta-l}\phi ^{\eta-l+1}\bigg]\label{eq:g_tau}
\end{align}
Colloquially, this expression can be read as the probability that the two direct contact nodes fail to connect the focal node to the GCC minus the probability that connection occurs through edges other than the two direct contact edges. 

The bond percolation properties of networks comprising weak cycles with fixed a degree of 4 is presented in Fig \ref{fig:WeakOutbreaks}. In the experiments, the networks are constructed of tree-like edges and a single type of weak cycle of increasing size using the generalised configuration model described in section REF. In the unclustered case the edges are all perfectly tree-like, reproducing exactly the configuration model; cycles form only through rare events, the probability of which vanishes in the limit of large and sparse graphs. Nodes in the clustered graphs are each part of two cycles; accidental joining of cycles to form higher-order structures is also a rare event. In each clustered graph, the joint degree sequence is therefore $p(k_\bot,k_\tau)=(0,2)$ whilst the configuration model has $p(k_\bot,k_\tau)=(4,0)$. We can see that clustering increases the critical point of the model; however, this effect is most distinguished for triangles. Higher-order weak cycles become increasingly tree-like locally and the importance of connection through the cycle diminishes. This can be understood from Eq \ref{eq:g_tau} where the contribution of probability to $g_\tau$ is reduced with increasing path length. 

\section{Higher-order strong clusters}
\label{sec:HOCLIQUES}

In this section we consider the percolation properties of strongly connected cycles for which there can be internal ``shortcuts'' across the cycle -- in particular fully connected sub-graphs, or cliques, of finite but unspecified order. In similar fashion to section \ref{sec:howc}, we index each node in the cycle from 1 to $\tau$ and set $\eta=\tau-1$. The probability that the first node sequentially attaches each node around the outside of the clique is 
\begin{equation}
    (1-u_\tau)(1-\phi)u_\tau^{\eta-1}\phi^\eta(1-\phi)^{\tau [(\tau-1)-2]/2}
\end{equation}
where $\tau [(\tau-1)-2]/2$ is the number of unique internal edges in the cycle. The next step is to evaluate the probability that a node in the body of the cycle connects the focal node to the giant component. Consider that the $i$th node in the cluster is connected to the giant component. In the first instance, we will compute the probability that it connects the focal node to the giant component through the outer path of the cycle. There must be a chain of unconnected nodes back to the focal node and the remainder of the cycle $i+1\ldots \eta$ must fail. The probability of success is
\begin{equation}
    (1-u_\tau)(1-\phi)u_\tau^{i-1}\phi^i 
\end{equation}
There are $1\ldots i-1$ nodes in the success-path. Each has connections to the focal node, other success path nodes, and failure-path nodes. The $j$th node in the success path must fail to attach any other node further down the chain, apart from its outer-skeleton neighbour, the $j-1$th node in the path. Each success path node therefore has the following failure probability
\begin{equation}
    (1-\phi)(1-\phi)^{j-2} 
\end{equation}
accounting for direct failure to connect the focal node and failure to attach the $j-2\ldots 1$ success-path nodes. We then account for each node in the chain from $i-1\ldots 2$ to arrive at the probability of success
\begin{equation}
    (1-u_\tau)(1-\phi)u_\tau^{i-1}\phi^i \prod^{i-1}_{j=2}(1-\phi)^{j-1}
\end{equation}
We do not consider node 1 as it must successfully attach the focal node. 
\begin{figure}[ht!]
\begin{center}
\includegraphics[width=0.475\textwidth]{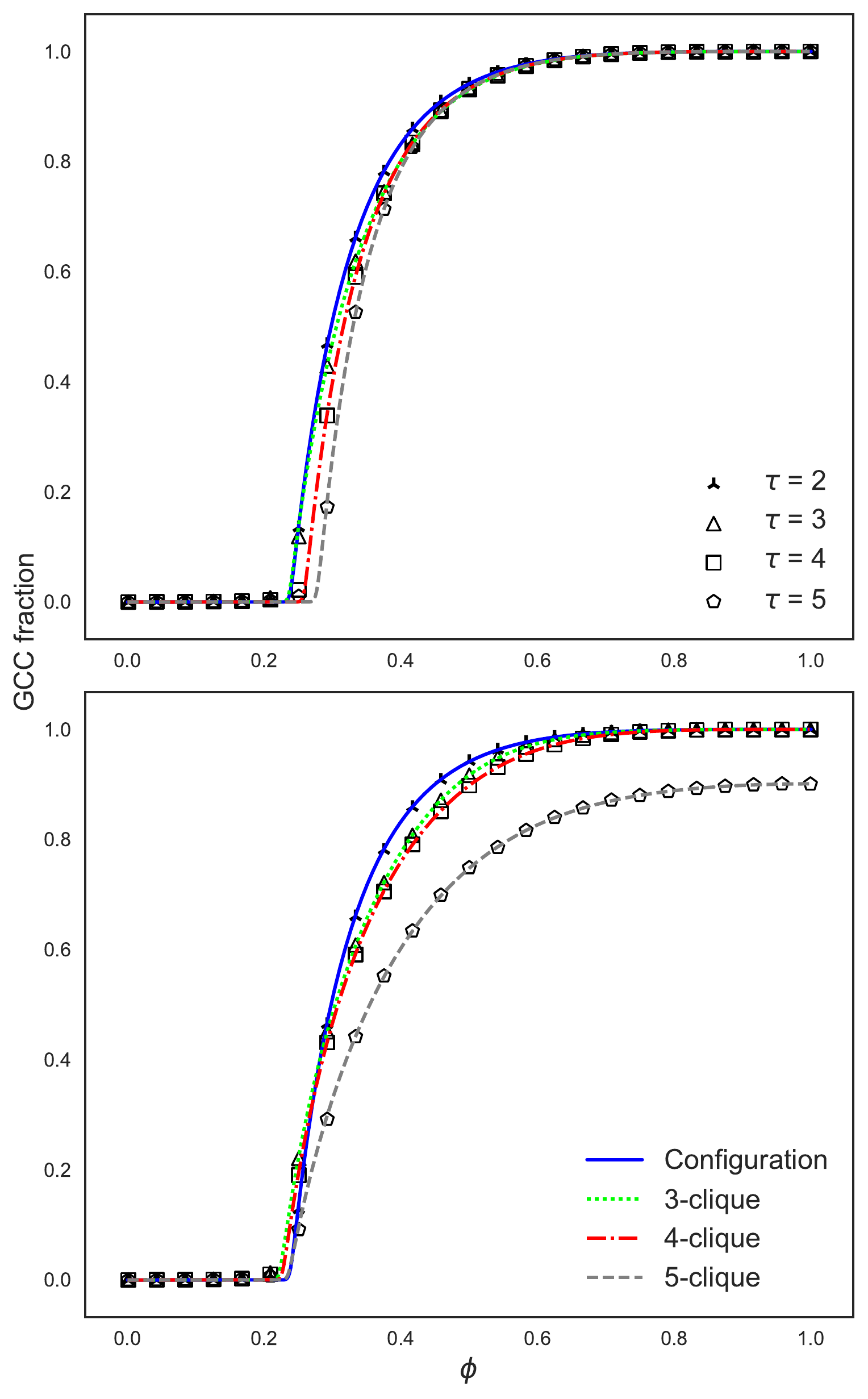}
\caption[network]{The fraction of the network occupied by the GCC as a function of bond occupancy for increasing clique sizes, $\tau$ with inverted assortativities. Nodes are either degree 4 or 6 with high (low) degree nodes tending to be clustered in the top (bottom) experiment. Markers indicate the experimental results while curves are the theoretical predictions from Eq \ref{eq:eqmainresclique}. From these experiments we observe that higher-order clustering increases the percolation threshold and reduces the size of the GCC. However, the dominant effect depends  on the assortativity of the clustering.
} \label{fig:StrongOutbreaks}
\end{center}
\end{figure}
Each node in the success-path also has connections to nodes in the failure path, indexed $i+1\ldots \eta$. There are $\eta-i$ failure path nodes. To evaluate these pathways for a given success-path node, we will consider 1 minus the success of each permissible pathway as $n$-hops through the failure-path nodes. This problem is equivalent to the enumeration of the non-self intersecting walks in a clique of dimension $(\eta-i+2)$, given that the start and end nodes are fixed for a particular success-path and focal node. A 1-hop walk starts on a given success-path node, traverses an edge to a failure path node and traverses another edge to the focal node with probability $u_\tau\phi^2$. There are $(\eta-i)$ of these walks, one for each failure path node. A 2-hop walk starts at a success-path node, hops to two failure-path sites and then finishes at the focal node with probability $u_\tau^2\phi^3$. There are $(\eta-i)$ failure-path nodes to choose from for the first hop and $(\eta-i-1)$ for the second, resulting in $(\eta-i)(\eta-i-1)$ potential pathways.
We have 
\begin{equation}
    [1-u_\tau\phi^2]^{(\eta-i)}[1-u_\tau^2\phi^3]^{(\eta-i)(\eta-i-1)}\ \cdots
\end{equation}
Given that there are $\eta-i$ failure-path nodes, this is the maximum non-self intersecting walk length and we have the total probability that a particular success-path node fails to connect the focal node to the giant component as
\begin{equation}
    \prod _{l=1}^{\eta-i}[1-u_\tau^l\phi^{l+1}]^{(\eta-i)_l}
\end{equation}
where
\begin{equation}
   (\eta-i)_l =  \prod^l_{k=1}(\eta-i-k+1)
\end{equation}
We must also account for the failure of each node in the failure-path to connect to the focal node. We do this by enumerating all $n$-hop walks that the $i+1\ldots \eta$ failure nodes can make. Following the same logic as for the success-path nodes we have 
\begin{equation}
    \bigg[\prod_{m=0}^{\eta-i-1} [1-u_\tau^m\phi^{m+1}(1-u_\tau)]^{(\eta-i)_m}\bigg]^{\eta-i}
\end{equation}
with the brackets representing the probability of failure of a single failure-path node, raised to the power of their occurrence, $\eta-i$. 

Finally, we account for the failure of the $i$th node via every other pathway apart from the successful path as
\begin{equation}
    \prod^{\eta-i}_{s=1}[1-u_\tau^s\phi^{s+1}]^{(\eta-i)_s}
\end{equation}

An important consequence of the symmetry of the clique is that paths of equal length through the cluster have equal probability of occurrence. Hence, while we derived the probability that the $i$th interior node traverses the outer skeleton of the clique, we are aware that this probability applies to all paths of length $i$ in the clique. Therefore, if we multiply this probability by the total number of walks of length $i$ in the cycle, we can account for all non-self intersecting walks that lead to the attachment of the focal node to the giant percolating component of the network. Since there are $\eta$ nodes apart from the focal node in the cycle the number of paths of length $i$ or the path-multiplicity is 
\begin{equation}
     \prod^i_{k=1}(\eta-k) = \begin{pmatrix}
    \eta-1\\ i
    \end{pmatrix} i!
\end{equation}
All that remains is to combine the elements of the probability and sum over all the potential path lengths in the clique and we arrive at the expression for $g_\tau$ that describes the probability that a node within a clique of size $\tau$ does not become attached to the giant component
\begin{widetext}
\begin{align}
    g_\tau =\ & [u_\tau+(1-u_\tau)(1-\phi)]^{\eta} - \eta!(1-u_\tau)(1-\phi) u_\tau^{\eta-1}\phi^\eta(1-\phi)^{\tau[(\tau-1)-2]/2}\nonumber\\
    &-\eta\sum^{\eta-1}_{i=2}\begin{pmatrix}
    \eta-1\\ i
    \end{pmatrix} i!(1-u_\tau)u_\tau^{i-1}\phi^i\prod^{i}_{j=2}(1-\phi)^{j-1}\prod^{\eta-i}_{l=1}[1-u_\tau^l\phi^{l+1}]^{(\eta-i)_l}\bigg\{\prod^{\eta-i-1}_{m=0}[1-u_\tau^m\phi^{m+1}(1-u_\tau)]^{(\eta-i)_m}\bigg\}^{\eta-i}\nonumber\\
    &\times \prod^{\eta-i}_{s=1}[1-u_\tau^s\phi^{s+1}]^{(\eta-i)_s}\label{eq:eqmainresclique}
\end{align}
\end{widetext}
To understand this formula in more detail we explicitly enumerate the probability, $g_{\diamondnew}$ that a node in a 4-clique does not become part of the GCC through edges in the motif in Appendix A. 

The bond percolation properties of networks comprising clique cycles is presented in Fig \ref{fig:StrongOutbreaks}. The experiments distribute the nodes evenly as either degree 4 or 6 but their clustering is distributed differently. In the first experiment (top) nodes of high degree tend to be clustered and contact other high degree nodes. The configuration model is generated by $p(4,0)=0.5$ and $p(6,0)=0.5$. For clique size $\tau=3$ we have $p(4,0)=0.5$, $p(2,2)=0.25$ and $p(0,3)=0.25$; $\tau=4$ we set $p(4,0)=0.5$, $p(3,1)=0.25$ and $p(0,2)=0.25$; and finally when $\tau=5$ we have $p(4,0)=0.5$ and $p(2,1)=0.5$. The local clustering coefficient, $c$, of each node-type in each network is given by $c=(0,0)$ for the configuration model; $c=(0,4/3,0.2)$ for the 4 degree and two 6 degree sites for the 3-clique network; $c=(0,0.2,0.4)$ for the 4 and two 6 degree sites when $\tau=4$; and, $c=(0,0.4)$ for the 4 and 6 degree sites when $\tau=5$. Whilst the degrees have been held constant in this trial, the local clustering coefficient is increasing with increasing $\tau$ due to the stronger clustering associated with higher-order cliques. It is clear that higher-order clustering is increasing the percolation threshold. 

The second experiment (Fig \ref{fig:StrongOutbreaks} bottom) is an inversion of the first: clustering now occurs predominantly within the low-degree sites and the degree 6 sites comprise mainly tree-like edges. The joint degree sequence is given by $p(4,0)=0.5$ and $p=(6,0)$ for the configuration model; $p(0,2)=0.25$, $p(2,1)=0.25$, $p(6,0)=0.5$ for $\tau=3$; $p(1,1)=0.5$, $p(6,0)=0.5$ for $\tau=4$ and finally $p(0,1)=0.5$, $p(2,1)=0.25$ and $p(6,0)=0.25$ for $\tau=5$. The local clustering coefficient of each site is $c=(0,0)$ for the configuration model; $c=(1/3, 1/6, 0)$ for $\tau=3$; $c=(1/2,0)$ for $\tau=4$ and $c=(1,0.4,0)$ for $\tau=5$. It is clear that as cliques become larger, the fraction of the network occupied by the GCC decreases for a given $\phi$. The increased probability for unclustered  high-degree nodes to contact one another increases their assortativity leading to a decrease in the epidemic threshold.

\begin{figure}[ht!]
\begin{center}
\includegraphics[width=0.475\textwidth]{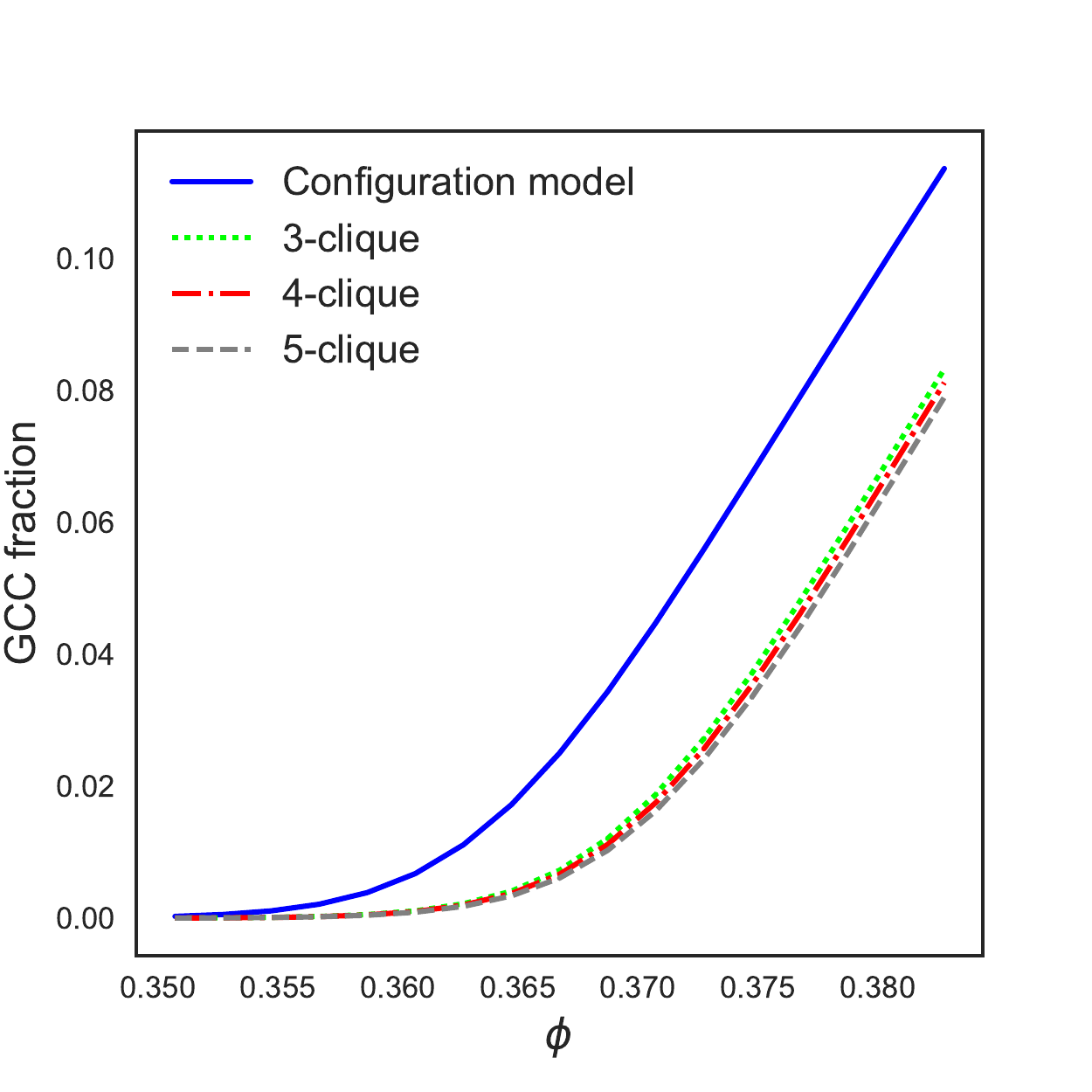}
\caption[network]{The fraction of the network occupied by the GCC as a function of bond occupancy for fixed degree and fixed local clustering coefficient, thus eliminating degree assortativity. It is clear that the increasing failure to be locally tree-like of larger cliques raises the critical threshold. 
} \label{fig:fixedstrong}
\end{center}
\end{figure}

In a final experiment, we set the degree of each cycle to $k=12$ in addition to constraining the local clustering coefficient across the clique sizes to be $c\approx 0.09$, thus eliminating assortativity. The degree distributions are given by: $p(12,0)=1$ for the configuration model; $p(0,6)=1$ for $\tau=3$; $p(6,2)=1$ for $\tau=4$ and $p(8,1)=1$ for $\tau=5$. It is clear from Fig \ref{fig:fixedstrong} that the percolation threshold increases with clique size despite each node belonging to a fixed number of triangles across the experiments. This is due to the additional correlations between first-order neighbours for the higher-order cycles.  

While in this section we have considered the non-self intersecting walks in cliques, it is hopefully clear that a hierarchy of cycles can be studied in the same way. In particular, all cycles whose nodes are degree-equivalent to one another, having $\tau$-fold rotational symmetry admit an analytical expression similar to the one presented here.  

\section{Exactness of analytical formulation}
\label{sec:exactness}

In the previous two sections we have developed an analytical formulation to compute the probability, $g_\tau$, that a node does not become attached to the GCC through its involvement in a cycle of length $\tau$. In this section, we show that this formulation is an approximation, albeit a very accurate one. 

To do this we compare the formulation for weak cycles (Eq \ref{eq:g_tau}) against a similar, exact, counting scheme. To develop the exact enumeration of states we must examine all configurations in a weak cycle. With $u_\tau$ taking its definition as before, we find
\begin{align}
    g^{\text{exact}}_\tau =\  &  (1-\phi)^2 + \sum^{\eta-1}_{i=1}(i+1)[\phi u_\tau]^i(1-\phi)^2\nonumber\\
    &+\tau[\phi u_\tau]^{\eta}(1-\phi)+[\phi u_\tau]^{\eta}\phi
    \label{eq:exact}
\end{align}
The first term considers the isolated focal node; the second accounts for the path on each side of the focal node being stopped by an unoccupied edge; only one edge is missing in the third term; and the fourth term considers the complete connected cycle. In other words, we have raised $u_\tau$ to the power of the number of nodes other than the focal node in the cycle for which there exists an occupied path linking them to the focal node. 

\begin{figure}[ht!]
\begin{center}
\includegraphics[width=0.49\textwidth]{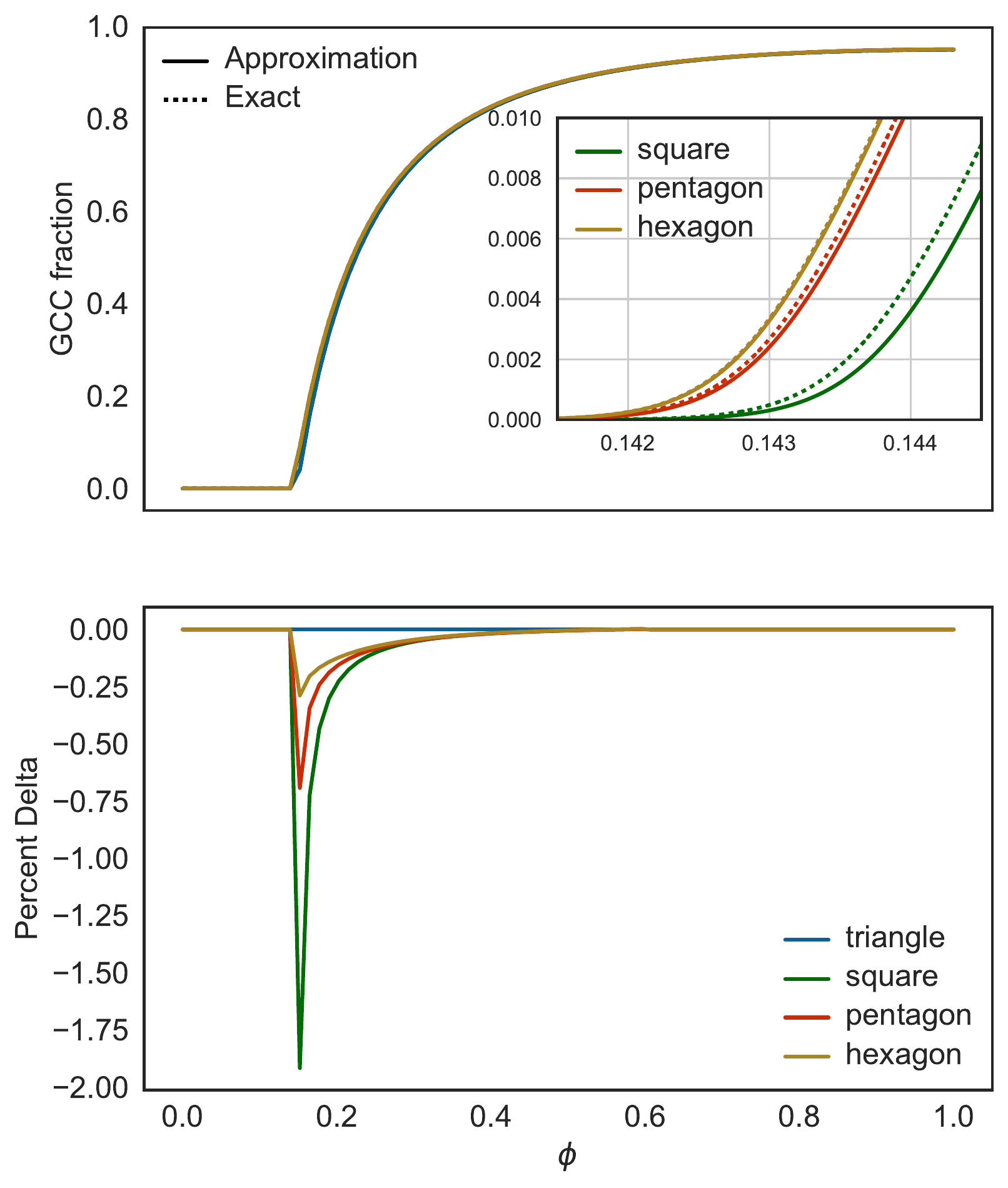}
\caption[EvA]{The fraction of the network occupied by the GCC for a Poisson distributed degree sequence against bond occupancy probability $\phi$ for the exact and approximate analytical solutions for a series of weak cycle topologies (top). The inset figure in the top tile shows a magnified section of this plot around the phase transition. Also plotted is the percentage difference between the approximate $A$ and exact $E$ solutions where $\delta=A-E$. We can see that while the difference is minor, it increases at the phase transition to a non-trivial fraction. Further, the extent of the approximation becomes vanishingly smaller as the cluster size increases. 
} \label{fig:ExactvsApprox}
\end{center}
\end{figure}

The results of a numerical investigation for networks with a Poisson degree sequence are depicted in Fig \ref{fig:ExactvsApprox}. We can see that the approximation breaks down at the onset of the GCC in the network. The approximation becomes increasingly more accurate following the phase transition for each topology, eventually coinciding once the GCC is established. This indicates that there is a missing probability that has not been accounted for by Eq \ref{eq:g_tau}, as the approximation slightly but consistently under-predicts the size of the GCC at the phase transition. 

On further investigation we notice that in the particular scenario whereby the entire cycle is connected by occupied edges the largest exponent of $u_\tau$ should be $\eta$, a case which is not represented by Eq \ref{eq:g_tau}. This term would become vanishingly small as $\phi\rightarrow 1$, but would be significant when the bond occupancy probability is critical -- \textit{i.e.}, precisely at the phase transition. Therefore it seems likely that there is missing mode of order $O(u_\tau^\eta)$ in Eq \ref{eq:g_tau} concerns the specific case in which all nodes within the cycle are not attached to the GCC. 

We cannot conduct an analysis for cliques or other site-equivalent cycles as currently there is no exact general equation for these cycles covering all orders. However, we can compare our expression (Eq \ref{eq:eqmainresclique}) for 4-cliques against the equation presented in Table 1 of \cite{karrer_newman_2010} (reproduced below for completeness). 
\begin{align}
    g_{\diamondnew}^{\text{exact}} =& (1-\phi)^3 +3\phi(1-\phi(2-\phi))^2u_{\diamondnew}\nonumber\\
    &
    +3\phi^2(3-2\phi)(1-3\phi+3\phi^2-\phi^3)u_{\diamondnew}^2\nonumber\\
    &+ \phi^3(16-33\phi+24\phi^2-6\phi^3)u_{\diamondnew}^3\label{eq:newman4clique}
\end{align}
We leave the unpacking and rationalisation of Eq \ref{eq:eqmainresclique} for 4-cliques to Appendix~\ref{sec:appendixA}.

\begin{figure}[ht!]
\begin{center}
\includegraphics[width=0.47\textwidth]{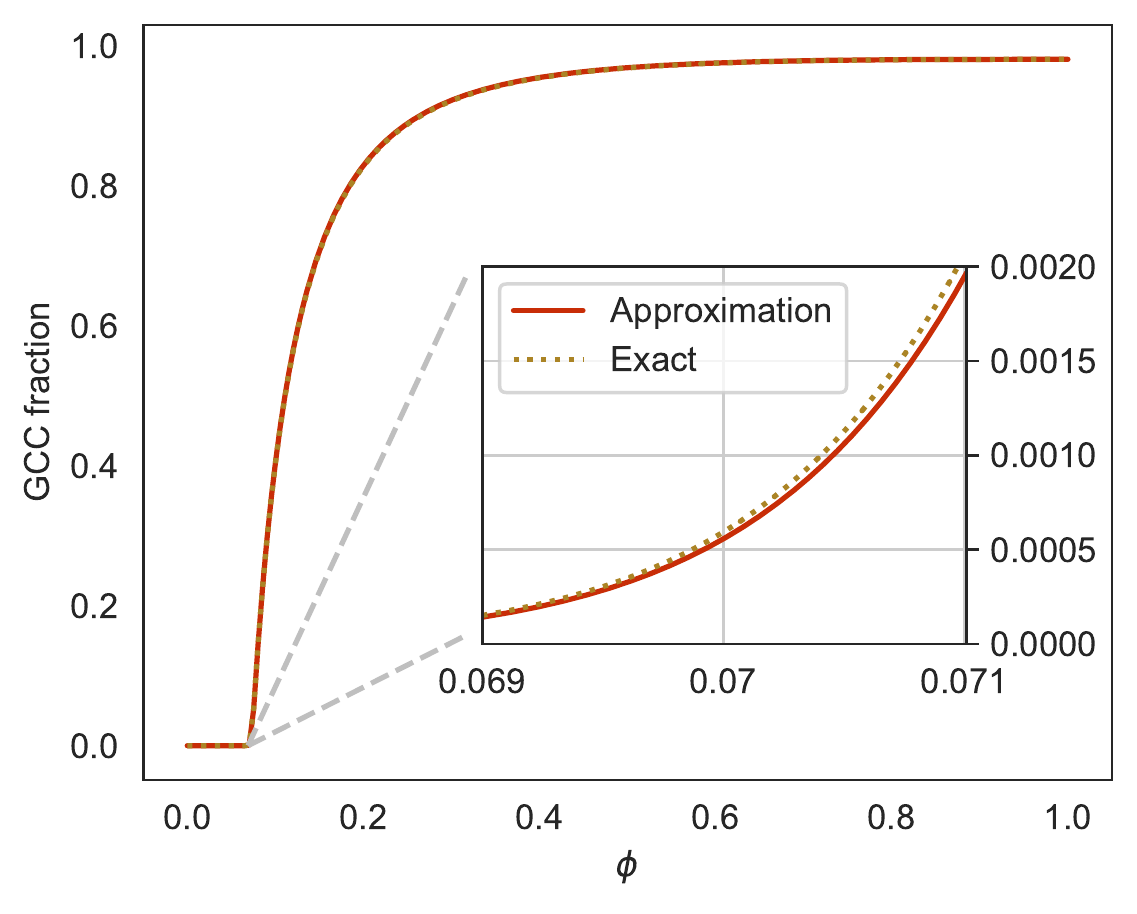}
\caption[EvAC]{The fraction of the network occupied by the GCC for a Poisson distributed degree sequence against bond occupancy probability $\phi$ for the exact (Eq \ref{eq:newman4clique}) and approximate (Eq \ref{eq:my4clique}) analytical solutions for the 4-clique graph. The inset figure shows a magnified section of this plot around the phase transition: note the scale of the inset compared to that of the main figure, showing how close the approximation is. Similarly to the weak-cycle case, we find our formula for the 4-clique graph to be in close agreement with the exact result presented by Karrer and Newman \cite{karrer_newman_2010}. 
} \label{fig:ExactvsApprox2}
\end{center}
\end{figure}

We find again that our formulation is an approximation to the exact result (see Fig \ref{fig:ExactvsApprox2}).
Therefore we conclude that, while our result is undoubtedly still an approximation, it is nonetheless in agreement with both the numerical results and the exact results (for the special case for which such a formulation exists) to a higher-order term of order $O(u_\tau^\eta)$.

\section{Mean component size and Percolation threshold}
\label{sec:Percolation threshold}

We can now calculate the sizes of the small components in the graph following the bond percolation process by defining $H_{1,\tau}(\bm z)$ as the generating function of the distribution of the number of vertices accessible from a node reached by following a random edge associated to a clique of size $\tau$ (a $\tau$-cycle). 
Following~\cite{newman2001rga, PhysRevLett.103.058701} this has a self-consistency solution
\begin{equation}
    H_{1,\tau}(\bm z) = zG_{1,\tau}(H_{1,\bot}(\bm z), \dots, H_{1,\gamma }(\bm z)) \label{eq:Newman1}
\end{equation}
for a given vector of topologies $\bm \tau = \{\bot,\Delta,\dots,\gamma \}$.

\noindent The probability that a node chosen at random belongs to a component of a given size is 
\begin{equation}
    H_{0}(\bm z) = z G_{0}(H_{1,\bot}(\bm z), \dots, H_{1,\gamma }(\bm z)) \label{eq:Newman2}
\end{equation}
\noindent The expectation value for the average component size in the network is then found by taking the derivative at $z=1$
\begin{equation}
    \langle H_{0} \rangle =G_{0}(H_{1,\bm{\tau}}(1)) + z\sum^\infty_{\nu\in {\bm \tau}}\frac{\partial G_{0} }{\partial H_{1,\nu}}\frac{\partial H_{1,\nu}}{\partial z}\bigg|_{z=1}\label{eq:averageCompSize}
\end{equation}
\noindent The derivatives $\partial _zH_{1,\bm{\tau}}(1)$ can be evaluated from equation~\ref{eq:Newman1}

\begin{equation}
\frac{\partial H_{1,\nu}}{\partial z}\bigg|_{z=1} = G_{1,\tau}(H_{1,{\bm\tau}}(1)) + z \sum^\infty_{\mu\in \bm\tau} \frac{\partial G_{1,\nu}}{\partial H_{1,\mu}}\frac{\partial H_{1,\mu}}{\partial z}\bigg|_{z=1}
\end{equation}
\noindent and with Eq. \ref{eq:Jacobian} we obtain a Hessian
\begin{align}
\frac{\partial H_{1,\nu}}{\partial z}\bigg|_{z=1} =\ & G_{1,\tau}(H_{1,{\bm\tau}}(1)) + z \sum^\infty_{\mu\in \bm\tau}\frac{\partial}{\partial H_{1,\mu}}\nonumber\\
&\cdot \bigg[\frac{1}{\langle k_\nu\rangle} \frac{\partial}{\partial H_{1,\nu}}G_{0}(H_{1,\nu}(1))\bigg]\frac{\partial H_{1,\mu}}{\partial z}\bigg|_{z=1}
\end{align}
\noindent This result can be rewritten as a matrix equation 
\begin{equation}
    \bm H = \bm 1 + {\alpha}^{-1}{H}\beta\cdot \bm H
\end{equation}
\noindent where $\bm 1 = (1,1, \dots)$ and $\bm H = (H_{1,\bot}, H_{1,\triangle},\dots )$ are vectors, $H$ is a Hessian of partial derivatives of $G_0(H_{1,\tau})$ with respect to $H_{1,\tau}$
\begin{equation}
    H = 
    \begin{pmatrix}
    \partial^2_{\bot,\bot} & \partial^2_{\bot,\triangle} & \dots &\partial^2_{\bot,\gamma } \\
    \partial ^2_{\triangle,\bot} & \partial ^2_{\triangle, \triangle}& \dots & \partial^2_{\triangle,\gamma }\\
    \vdots& \vdots & \ddots & \vdots\\
    \partial^2_{\gamma , \bot} & \partial^2_{\gamma , \triangle}& \dots &  \partial^2_{\gamma ,\gamma } 
    \end{pmatrix}
\end{equation}
\noindent 
where $\alpha$ is a diagonal matrix of expected values of the number of cycles of a given topology.
\begin{equation}
    \alpha = 
    \begin{pmatrix}
    \langle k_\bot\rangle & 0 & \dots  & 0 \\
    0 & \langle k_\triangle \rangle & \dots  & 0 \\
    \vdots & \vdots & \ddots & \vdots \\
    0 & 0 & \dots  & \langle k_\gamma \rangle
\end{pmatrix}
\end{equation}
\noindent and $\beta$ is given by
\begin{equation}
    \beta = 
    \begin{pmatrix}
    1 & 0 & \dots  & 0 \\
    0 & 2& \dots  & 0 \\
    \vdots & \vdots & \ddots & \vdots \\
    0 & 0 & \dots  & \gamma -1
\end{pmatrix}
\end{equation}
\noindent which is a diagonal matrix of the number of direct contacts the focal node has to the cycle. 

Rearranging this equation allows us to solve for the derivatives in Eq \ref{eq:averageCompSize} to find the average component size $(I - {\alpha}^{-1} H\beta)\cdot \bm H = \bm 1$ where $I$ is the identity matrix. When the determinant vanishes, $\det( I - {\alpha}^{-1} H\beta) = 0$, the average component size diverges, signalling the onset of the giant component. The determinant of a $\tau\times \tau$ matrix can be written using the $\tau$-dimensional Levi-Civita symbol
\begin{equation}
    \sum^{\tau}_{i_1=1}\sum^{\tau}_{i_2=1}\dots \sum^{\tau}_{i_\tau=1} \epsilon_{i_1\dots i_\tau} x_{1j_1}\dots x_{\tau j_\tau}
\end{equation}
\noindent where diagonal elements are given by $1- \partial^2_{\tau,\tau} \slash {\langle k_\tau\rangle}$ and off-diagonal elements are $-\partial^2_{\mu,\nu} \slash {\langle k_\mu\rangle}$. The appropriate generalisation of the Molloy-Reed criterion \cite{molloy_reed_1995} for networks containing cycles is then found by evaluating this expression. A GCC can be found in the network when 
\begin{equation}
    \det( I -{\alpha}^{-1} H\beta)\leq 0\label{eq:mat}
\end{equation}
For instance, when the network contains only tree-like edges, $\bm \tau=\{\bot\}$, then the determinant in Eq \ref{eq:mat} yields the familiar Molloy-Reed criterion, $\langle k^2_\bot\rangle/\langle k_\bot\rangle -2 = 0$. When the network consists of tree-like and triangular edges, $\bm \tau=\{\bot,\triangle\}$, Eq \ref{eq:mat} reduces to
\begin{equation}
    \left(\frac{\langle k_\bot^2\rangle - \langle k_\bot\rangle }{\langle k_\bot\rangle}-1 \right) \left(2\frac{\langle k_\triangle^2\rangle - \langle k_\triangle\rangle }{\langle k_\triangle\rangle}-1  \right)\leq 2 \frac{\langle k_\bot k_\triangle\rangle^2}{\langle k_\bot \rangle\langle k_\triangle\rangle}\label{eq:matMRtreeTrianglemod}
\end{equation}
a result obtained by \cite{miller_2009, PhysRevLett.103.058701}. If the network contained subgraphs of larger order, such as $\bm \tau = \{\bot , \triangle, \diamondnew\}$ where $\diamondnew$ is the 4-clique, then the condition for the onset of the GCC is given by 
\begin{widetext}
\begin{align}
      &\left[1-\frac{\langle {k^2_\bot}\rangle-\langle k_\bot\rangle }{\langle k_\bot\rangle} \right]\left[1-2\frac{\langle  {k^2_\triangle}\rangle-\langle  k_\triangle\rangle }{\langle  k_\triangle\rangle} \right]\left[1-3\frac{\langle  {k^2_{\diamondnew}}\rangle-\langle  {k_{\diamondnew}}\rangle }{\langle  {k_{\diamondnew}}\rangle} \right] 
    + \left[-2\frac{\langle k_\bot  k_\triangle\rangle}{\langle  k_\triangle\rangle} \right]\left[-3\frac{\langle  k_\triangle {k_{\diamondnew}}\rangle}{\langle { k_{\diamondnew}}\rangle} \right]\left[-\frac{\langle{ k_\bot  k_{\diamondnew}}\rangle}{\langle k_\bot \rangle} \right]\nonumber\\
  & +\left[-3\frac{\langle {k_\bot  k_{\diamondnew}}\rangle}{\langle  {k_{\diamondnew}}\rangle} \right]
    \left[-\frac{\langle k_\bot  k_\triangle\rangle}{\langle k_\bot \rangle} \right]
    \left[-2\frac{\langle  {k_\triangle k_{\diamondnew}}\rangle}{\langle  k_\triangle\rangle} \right]
    -\left[-3\frac{\langle {k_\bot  k_{\diamondnew}}\rangle}{\langle  {k_{\diamondnew}}\rangle} \right]
    \left[1-2\frac{\langle  {k^2_\triangle}\rangle-\langle  k_\triangle\rangle }{\langle  k_\triangle\rangle} \right]
    \left[-\frac{\langle {k_\bot  k_{\diamondnew}}\rangle}{\langle k_\bot \rangle} \right]\nonumber\\
    &-\left[-2\frac{\langle k_\bot  k_\triangle\rangle}{\langle  k_\triangle\rangle} \right]
    \left[-\frac{\langle k_\bot  k_\triangle\rangle}{\langle k_\bot \rangle} \right]
    \left[1-3\frac{\langle { k^2_{\diamondnew}}\rangle-\langle  {k_{\diamondnew}}\rangle }{\langle  {k_{\diamondnew}}\rangle} \right]
    -\left[1-\frac{\langle k^2_\bot \rangle-\langle k_\bot \rangle }{\langle k_\bot \rangle} \right]
    \left[-3\frac{\langle  {k_\triangle k_{\diamondnew}}\rangle}{\langle  {k_{\diamondnew}}\rangle} \right]
    \left[-2\frac{\langle  {k_\triangle k_{\diamondnew} }\rangle}{\langle {k_\triangle}\rangle} \right]\leq 0
\end{align}
\end{widetext}
It is clear that when the network contains no triangles and 4-cliques only the first bracket of the first term survives, which is the Molloy-Reed criterion for tree-like networks. Additionally, if only tree-like and triangle motifs are found, then we recover Eq \ref{eq:matMRtreeTrianglemod}.

For networks composed of a single clique-type of size $\tau$, the Molloy-Reed criterion is given by
\begin{equation}
    \left((\tau-1)\frac{\langle k_\tau^2\rangle}{\langle k_\tau\rangle}-\tau\right) 
    \leq 0
\end{equation}

\section{numerical examples}
\label{sec:numExample}

As a numerical example we assign each topological cycle a Poisson degree distribution, approximating the Erd\H{o}s-Renyi model for large network size. Since each each $k_\tau$ is an independent variable this is simply a product of independent Poisson distributions
\begin{equation}
    p(k_{\bm\tau}) = \prod_{\tau\in \bm\tau}e^{-\langle k_{\tau}\rangle}\frac{\langle k_{\tau}\rangle^{k_{\tau}}}{k_{\tau}!}\label{eq:Pois}
\end{equation}
\noindent where the product extends over each topology and $\langle k_\tau\rangle$ is the average number of tree-like edges, triangles, squares (and so on) per node.  This is generated using Eq. \ref{eq:G0}
\begin{equation}
    G_0(z) = \prod_{\tau\in \bm\tau}e^{\langle k_{\tau}\rangle(z-1)}
\end{equation}
\noindent since
\begin{equation}
    e^{z\langle k_\tau\rangle} = \sum_{k_\tau=0}^\infty \frac{(z\cdot\langle k_\tau\rangle)^{k_\tau}}{k_\tau!}
\end{equation}
It is clear that in this case $G_{1,\tau}(z)=G_1(z)$ and hence, the outbreak fraction satisfies the transcendental equation 
\begin{equation}
    S = 1 - e^{S}\prod_{\tau \in \bm \tau} e^{-\langle k_\tau\rangle}
\end{equation}
which can be solved by fixed-point iteration.  To see this, consider a network consisting of tree-like, 3- and 4-node cliques each with a Poisson distribution such that each node has an average of $(\mu,\nu,\eta)$ of each respective cycle. For $\phi=1$, $S$ is given by
\begin{equation}
S = 1-e^{-\mu S}e^{-\nu S(2-S)}e^{-\eta S(S^2-3S+3)}\label{eq:poisson example}
\end{equation}


We can also consider a power-law degree distribution with exponential degree cut-off of the form
\begin{equation}
    p(k_{\bm\tau}) =  C \prod_{\tau\in \bm\tau} k_\tau ^{-\alpha_\tau} e^{-k_\tau/\kappa_\tau}\label{eq:sfwc}
\end{equation}
where $C$, $\alpha_\tau$ and $\kappa_\tau$ are constants for $k\geq 1$ for the following set of degrees
\begin{equation}
    \{k_\tau \in \mathbb Z^+_0\ |\ k_\tau=0\text{ mod }(\tau-1)\}
\end{equation}
The normalisation constant can be found from the condition $G_0(1)=1$ 
\begin{equation}
    C^{-1} = \sum_{k_\bot=1}^\infty \sum_{k_\triangle=1}^\infty \cdots \sum_{k_\tau=1}^\infty \frac{e^{-k_\bot/\kappa_\bot}}{k^{\alpha_\bot}}\frac{e^{-k_\triangle/\kappa_\triangle}}{k^{\alpha_\triangle}}\cdots \frac{e^{-k_\tau/\kappa_\tau}}{k^{\alpha_\tau}}
\end{equation}
which is a multipolylogarithm or the form
\begin{equation}
    \text{Li}_{s_1,\dots, s_k}(z_1,\dots,z_k) = \sum_{n_1>\dots>n_k>0}\left(\prod^k_{j=1}n_j^{-s_j}z_j^{n_j}\right)
\end{equation}
which is convergent on the disc $|z_\tau|<1$ $\forall \tau$.
The $G_0(\bm z)$ and  $G_{1,\nu}(\bm z)$ generating functions can then be computed as 
\begin{widetext}
\begin{equation}
    G_0(z_\bot,\dots,z_\tau) = \frac{\text{Li}_{\alpha_\bot,\dots,\alpha_\tau}(z_\bot e^{-1/\kappa_\bot},\dots,z_\tau e^{-1/\kappa_\tau})}{ \text{Li}_{\alpha_\bot,\dots,\alpha_\tau}(e^{-1/\kappa_\bot},\dots,e^{-1/\kappa_\tau})}
\end{equation}

\begin{equation}
    G_{1,\nu}(z_\bot,\dots,z_\tau) = \frac{\text{Li}_{\alpha_\bot,\dots,\alpha_\nu-1,\dots,\alpha_\tau}(z_\bot e^{-1/\kappa_\bot},\dots,z_\tau e^{-1/\kappa_\tau})}{ z_\nu \text{Li}_{\alpha_\bot,\dots,\alpha_\nu-1,\dots,\alpha_\tau}(e^{-1/\kappa_\bot},\dots,e^{-1/\kappa_\tau})}
\end{equation}
\end{widetext}
and when $\kappa_\tau\rightarrow \infty$ $\forall \tau \in \bm \tau$ we have purely power-law networks. In this case we have the outbreak fraction given by Eq. \ref{EQ:main} where each $z_\nu$ value is computed as 
\begin{equation}
    z_\nu = \frac{\text{Li}_{\alpha_\bot,\dots,\alpha_{\nu}-1,\dots,\alpha_\tau}(z_\bot,\dots,z_\tau )}{z_\nu\zeta(\alpha_\bot,\dots,\alpha_{\nu-1},\dots,\alpha_{\tau})}
\end{equation}
where $\zeta(s_1,\dots,s_k)$ are the multiple Riemann-zeta values
\begin{equation}
    \zeta(s_1,\dots,s_k)=\sum_{n_1>\dots>n_k>0}\left(\prod^k_{j=1}n_j^{-s_j}\right)
\end{equation}

\section{Conclusion}
\label{sec:conclusion}
In this paper we have used the generating function formulation to consider networks containing higher-order clustering of both weak cycles and $\tau$-cliques. We have derived the size of the giant component, the mean component size and generalised the Molloy-Reed criterion describing the onset of the phase transition for these networks. We have provided analytical results for Poisson and power-law networks with exponential degree cut-off. We have shown that higher-order weak cycles behave increasingly more tree-like; the triangle introduces the greatest deviation from the independent edge solution. 
Conversely, cliques lead to a higher percolation threshold when considering the degree-$\delta$ model introduced by Miller. It was also shown that the final size of the GCC can be reduced by clustering, although this is accompanied by a reduction in the critical threshold due to assortativity of low-degree sites. We have also demonstrated that the emergent properties of complex networks are intricately determined by other factors, chief among these is the degree assortativity. In a final experiment, we held the degree and local clustering coefficient of the network constant. We found that higher-order cliques increase the percolation threshold; however, the magnitude of the effect diminishes with size. 

We have not discussed clusters of intermediate strength formed through the sequential weakening of the $\tau$-clique. These cycles contain more than one type of node in their skeleton, ruining the symmetry of the motif. And while the percolation properties of a particular cycle may be straightforward to evaluate, the generalisation to all order appears to be non-trivial. 

The approach presented here can readily be generalised to the study of directed, multi-layer and weighted networks. With the enumeration of flow avenues through a clustered network, one can imagine the construction of tailored disease control strategies that target prominent spreading pathways rather than only prominent nodes. This work also has implications for community detection \cite{krzakala_moore_mossel_neeman_sly_zdeborova_zhang_2013,martin_zhang_newman_2014}, as well as the study of the structural properties of empirical networks. 

\subsection{Acknowledgments}
The authors thank three anonymous referees for their useful comments and in particular referee 2 who provided equation \ref{eq:exact}. 

\subsection*{References}
\bibliography{CLU}

\appendix
\section{4-cliques}
\label{sec:appendixA}

In this appendix we will explicitly evaluate Eq \ref{eq:eqmainresclique} for the 4-clique, which we symbolise with $\tau = \;\diamondnew$. The key to understanding the formulation is that walks of a given length in a clique have equal probability of occurring. Therefore, we must count all walks of a given length through the clique from all potential source nodes to the focal node and then enumerate the probability of this mode. We find that the probability that a node involved in a 4-clique belonging to a finite-sized component during bond percolation is given by
\begin{widetext}
\begin{equation}
    g_{\diamondnew} =(u_{\diamondnew}+(1-u_{\diamondnew})(1-\phi) )^3-6(1-u_{\diamondnew})u_{\diamondnew}\phi ^2(1-\phi )(1-u_{\diamondnew}\phi ^2)^2(1-(1-u_{\diamondnew})\phi )-6(1-u_{\diamondnew})u_{\diamondnew}^2\phi ^3(1-\phi )^3\label{eq:my4clique}
\end{equation}
\end{widetext}
The rationalisation of this expression is quite simple and can be read from left to right as follows. Consider a 4-clique and choose a node to be the focal node. The  first cubic term relates to the failure of the three direct-contact nodes to connect the focal node to the GCC. These are 0-hop walks as they concern the direct linkage to the focal node. 
\begin{figure}[ht!]
\begin{center}
\includegraphics[width=0.4\textwidth]{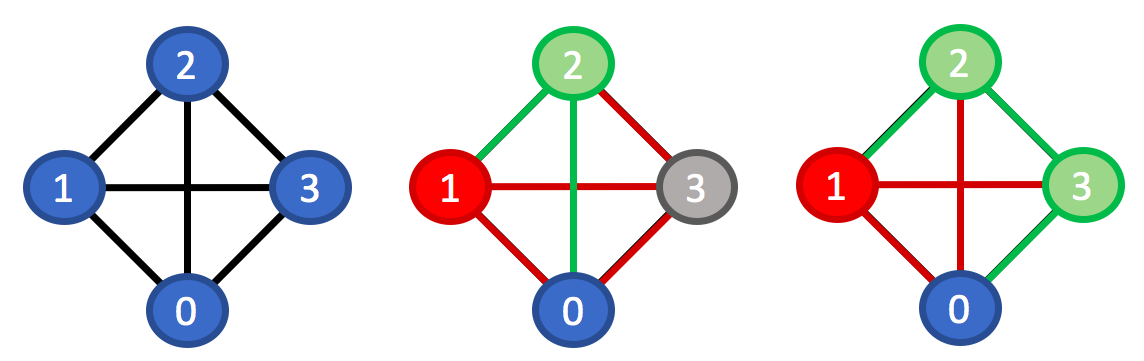}
\caption[4cliqueexample]{ 
The 4-clique (left) with labelled node sites and focal node chosen to be node 0. Assuming that node 1 is attached to the GCC (red node), then there are two types of non-direct walks back to the focal node. The 1-hop walk (center) requires that node 2 is not attached to the GCC (green). Bond occupation must occur through the path [1,2,0], which we term the success path. The state of node 3 is unspecified by the success path (grey). However, all other paths, from any starting node, that does not cause intersection with the success path, must fail to attach node 0 to the GCC (red edges).  We term these the failure paths for the given success path under consideration. For the center success path, the failure paths are [1,3,0], [2,3,0] and [3,0]. The first two assume that node 3 is in state $u_{\diamondnew}$, while the final path assumes node 3 was attached to the GCC prior to this. 
} \label{fig:4cliqueexample}
\end{center}
\end{figure}
Labeling the nodes according to Fig \ref{fig:4cliqueexample} (left) we notice that if node 1 fails to connect the focal node directly, it can still connect it through edges in the clique. There are two distinct paths that can be made back to the focal node: 1-hop (center) and 2-hop (right) walks. 

The second term in Eq \ref{eq:my4clique} concerns the 1-hop walks in the clique. Consider (for instance) that node 1 is the source node. For this walk to occur node 1 must be attached to the GCC with probability $(1-u_{\diamondnew})$, but it has failed to attach the focal node directly with probability $1-\phi$. Node 2 (for instance) must become attached through bond occupation from node 1 with probability $u_{\diamondnew}\phi$, which then goes on to connect to the focal node through its direct edge with probability $\phi$. We then must ensure that all the remaining pieces in the clique that have not been assigned a probability must be dealt with, we cannot leave them unaccounted for. Both node 1 and node 2 must fail to exercise their alternative 1-hop walks back to the focal node. The probability of each of these walks failing is $1-u_{\diamondnew}\phi^2$. However, it might happen that node 3 was also attached to the GCC, in which case, it must fail directly with probability $(1-u_{\diamondnew})(1-\phi)$. The factor of 6 accounts for the path multiplicity; each node has 2 1-hop walks back to the focal node. For instance, we depicted the success path in Fig \ref{fig:4cliqueexample} as [1,2,0], however, another valid 1-hop walk from 1 is [1,3,0]. 

The final term is much easier to rationalise. Consider again that node 1 is attached to the GCC, but that it has failed directly to connect to the focal node, Fig \ref{fig:4cliqueexample}. The 2-hop walk [0,2,3,0] back to the focal node around the clique must fix both nodes 2 and 3 to be unattached and involve three bond occupation events. Further, both interior edges in the clique must not short-circuit the 2-hop walk into a 1-hop walk, so they too, in addition to node 1s direct edge, must be unoccupied. The other 2-hop walk starting from 1 is given by [1,3,2,0].

\end{document}